\documentclass[twocolumn,floatfix]{aastex631}

\newcommand{\hb}{H$\beta$}

\newcommand{\ho}{$H_0$}

\newcommand{\kmsm}{kms$^{-1}$Mpc$^{-1}$}

\newcommand{\Msol}{M$_\odot$}
\newcommand{\Mstar}{M\textsubscript{*}}
\newcommand{\tilda}{$\sim$}
\newcommand{\deltaM}{$|$$\Delta$\Mstar$|$}
\newcommand{\deltaz}{$|$$\Delta$z$|$}
\newcommand{\OIIIHb}{[OIII]/\hb}
\newcommand{\lessthan}{$<$}
\newcommand{\greaterthan}{$>$}

\graphicspath{{./}{figures/}}

\begin{document}


\title{Exposing Line Emission: A First Look At The Systematic Differences of Measuring Stellar Masses With JWST NIRCam Medium Versus Wide Band Photometry}


\author[0000-0001-8830-2166]{Ghassan T. E. Sarrouh}
\affiliation{Department of Physics and Astronomy, York University, 4700 Keele St., Toronto, Ontario MJ3 1P3, Canada}

\author[0000-0002-9330-9108]{Adam Muzzin}
\affiliation{Department of Physics and Astronomy, York University, 4700 Keele St., Toronto, Ontario MJ3 1P3, Canada}


\author[0000-0001-9298-3523]{Kartheik G. Iyer}
\affiliation{Columbia Astrophysics Laboratory, Columbia University, 550 West 120th Street, New York, NY 10027, USA}

\author[0000-0002-8530-9765]{Lamiya Mowla}
\affiliation{Whitin Observatory, Department of Physics and Astronomy, Wellesley College, 106 Central Street, Wellesley, MA 02481, USA}

\author[0000-0002-4542-921X]{Roberto Abraham}
\affiliation{David A. Dunlap Department of Astronomy and Astrophysics, University of Toronto, 50 St. George Street, Toronto, Ontario, M5S 3H4, Canada}

\author[0000-0003-3983-5438]{Yoshihisa Asada}
\affiliation{Department of Astronomy and Physics and Institute for Computational Astrophysics, Saint Mary's University, 923 Robie Street, Halifax, Nova Scotia B3H 3C3, Canada}
\affiliation{Department of Astronomy, Kyoto University, Sakyo-ku, Kyoto 606-8502, Japan}

\author[0000-0001-5984-0395]{Maru\v{s}a Brada{\v c}}
\affiliation{University of Ljubljana, Department of Mathematics and Physics, Jadranska ulica 19, SI-1000 Ljubljana, Slovenia}
\affiliation{Department of Physics and Astronomy, University of California Davis, 1 Shields Avenue, Davis, CA 95616, USA}

\author[0000-0003-2680-005X]{Gabriel B. Brammer}
\affiliation{Cosmic Dawn Center (DAWN), Denmark}
\affiliation{Niels Bohr Institute, University of Copenhagen, Jagtvej 128, DK-2200 Copenhagen N, Denmark}

\author[0000-0001-8325-1742]{Guillaume Desprez}
\affiliation{
Department of Astronomy \& Physics and Institute for Computational Astrophysics, Saint Mary's University, 923 Robie Street, Halifax, Nova Scotia, B3H 3C3, Canada
}


\author[0000-0003-3243-9969]{Nicholas S. Martis}
\affiliation{University of Ljubljana, Department of Mathematics and Physics, Jadranska ulica 19, SI-1000 Ljubljana, Slovenia}

\author[0000-0002-7547-3385]{Jasleen Matharu}
\affiliation{Cosmic Dawn Center (DAWN), Denmark}
\affiliation{Niels Bohr Institute, University of Copenhagen, Jagtvej 128, DK-2200 Copenhagen N, Denmark}

\author{Gaël Noirot}
\affiliation{Department of Astronomy and Physics and Institute for Computational Astrophysics, Saint Mary's University, 923 Robie Street, Halifax, Nova Scotia B3H 3C3, Canada}

\author[0000-0002-7712-7857]{Marcin Sawicki}
\affiliation{Department of Astronomy and Physics and Institute for Computational Astrophysics, Saint Mary's University, 923 Robie Street, Halifax, Nova Scotia B3H 3C3, Canada}

\author[0000-0002-6338-7295]{Victoria Strait}
\affiliation{Cosmic Dawn Center (DAWN), Denmark}
\affiliation{Niels Bohr Institute, University of Copenhagen, Jagtvej 128, DK-2200 Copenhagen N, Denmark}

\author[0000-0002-4201-7367]{Chris J. Willott}
\affiliation{NRC Herzberg, 5071 West Saanich Rd, Victoria, BC V9E 2E7, Canada}

\author{Johannes Zabl}
\affiliation{Department of Astronomy and Physics and Institute for Computational Astrophysics, Saint Mary's University, 923 Robie Street, Halifax, Nova Scotia B3H 3C3, Canada}


\begin{abstract}

Photometrically derived stellar masses are known to suffer from systematic uncertainties, particularly due to nebular emission contributions to the spectral energy distribution. Using \emph{JWST} NIRCam imaging from the CAnadian NIRISS Unbiased Cluster Survey (CANUCS), we introduce a comparison study of photometrically-derived redshifts and stellar masses based on two photometric catalogs of the same field spanning $\sim$0.4-4.5$\mu$m: one consisting solely of wide band photometry, and another employing a combination of wide and medium band photometry. 
We find that \tilda70\% of galaxies have consistent photometric redshifts between both catalogs, with median stellar mass difference between the two catalogs of \lessthan\ 0.2 dex across all redshift bins. There are however a subset of galaxies (5\% at z\tilda2 up to 15\% at z\tilda6) where wide bands underestimate star formation rates and infer older stellar populations, leading to median stellar mass differences of \tilda0.7 dex. Examination of the SEDs for galaxies with inconsistent photometric redshifts shows this is caused by the inability of the wide bands to distinguish continuum emission from emission lines.  Computing a stellar mass density with our sample we find that it is potentially underestimated using wide-band photometry by \tilda10-20\% at z \lessthan\ \ 4, and potentially overestimated by as much as a factor of 2-3 at z \greaterthan\ 5. These systematic differences caused by the poor spectral resolution of wide bands have implications for both ongoing and future planned observing programs which determine stellar mass and other physical properties of high redshift galaxies solely via wide band photometry.

\end{abstract}


\section{Introduction} \label{sec:intro}

Photometric surveys are instrumental to generating large statistical samples of galaxies owing to their efficiency compared to obtaining spectra. Photometric redshift and spectral energy distribution (SED) fitting has become a primary tool in constraining galaxy properties \cite{Loh1986}, with numerous codes developed to convert photometric measurements into redshifts and physical properties (e.g. \citealt{Brammer2008, Conroy2009, Kriek2009, Carnall2019,Iyer2019, Johnson2021}).
To efficiently collect data where spectroscopy is either impractical or infeasible, wide band photometry has been used ubiquitously throughout space-based and ground-based surveys (e.g. \citealt{Koekemoer2011, McCracken2012, Lotz2017}) and is in continued use with the James Webb Space Telescope (\emph{JWST}) owing to its wide spectral coverage and broad sampling of the SED.

Key metrics of galaxy evolution (e.g. the stellar mass function, the size-mass relation, the star formation main sequence) typically rely on galaxy properties derived from wide band photometry (see \citealt{Robertson2021} for a review). However, previous ground-based optical medium band surveys (e.g. SHARDS, \citealt{Perez-Gonzalez2013}; COMBO-17, \citealt{Wolf2003}) have shown the inefficacy of wide band photometry to constrain high-redshift SEDs in the presence of strong emission or absorption lines.

Accurately modelling the strong nebular emission seen by \emph{JWST} in high redshift spectra (e.g. \citealt{Withers2023, Boyett2022}) is critical to accurately constraining galaxy properties of photometrically-derived galaxy samples. 
Previous work has shown that nebular emission in the rest-frame optical region of SEDs at z \greaterthan \ 2 contaminates wide band photometric measurements, biasing physical properties of galaxies at early times (\citealt{Stark2013, Smit2014}). Nevertheless, the largest ERS programs by area (CEERS, \citealt{FinkelsteinCEERS2017proposal}) and depth (GLASS, \citealt{Treu2022}), as well as some of the largest Cycle 1 programs in legacy fields (COSMOS-Web,  \citealt{Casey2023}; JADES in GOODS-N/GOODS-S,  \citealt{Eisenstein2023}) only use up to two medium bands. JEMS \citep{Williams2023}, a public Cycle 1 program in the Hubble Ultra Deep Field, offers deep medium band imaging at \tilda2$\mu$m \& \tilda4.5$\mu$m in 5 filters. However the area of the survey is limited to a single NIRCam pointing (\tilda 10 arcmin$^{2}$), and the lack of continuous coverage across the near infrared spectrum means the data set is only sensitive to particular spectral features at particular redshifts. Multiple approved Cycle 2 programs are emphasizing medium band photometry, including large follow up programs in the JADES \citep{Eisenstein2023a}, UNCOVER (ID: 4111, PI: K. Suess), and CANUCS (ID: 3362, PI: A. Muzzin) fields. Fields with a combination of wide and medium band photometry will be instrumental in exploring systematic effects between these two sets of measurements more fully and understanding their relative utility.

\cite{Finkelstein2022} have shown that template fitting codes such as \texttt{EAZY} \citep{Brammer2008} lack sufficiently blue templates to reflect the highly star forming galaxies prevalent at high redshifts, leading to the development of templates incorporating stronger emission lines with \texttt{CLOUDY} \citep{Ferland2017}, binary stellar evolution with \texttt{BPASS} \citep{Stanway2018} and younger galaxy populations to better capture the expected bluer rest-frame UV color of such sources \citep{Larson2023}. Resolving these spectral features is thus imperative for ensuring the fidelty of photometrically-derived data sets and the accuracy of the science that results. The intention of this study is to introduce a comparative analysis between photometric data sets which include or exclude medium band photometry. Taking a field with overlapping wide and medium band imaging, we construct a purely wide band photometric data set, and a second set which also includes all available medium bands. The wide band photometry can then be used as a baseline against which to compare differences in derived properties, such as redshift or stellar mass as in this work, resulting from the inclusion of medium band photometry.

This work presents a comparison of \emph{JWST} NIRCam wide and medium band imaging, described in \S \ref{sec:data}. Photometry, redshift/SED-fitting, and methodology is discussed in \S \ref{sec:methodology}. Results are present in \S \ref{sec:results}. This work assumes a cosmology of \ho = 70\kmsm, $\Omega$$_{m}$ = 0.30, and $\Omega$$_{\Lambda}$ = 0.70.

\begin{figure*}[t!]
\plotone{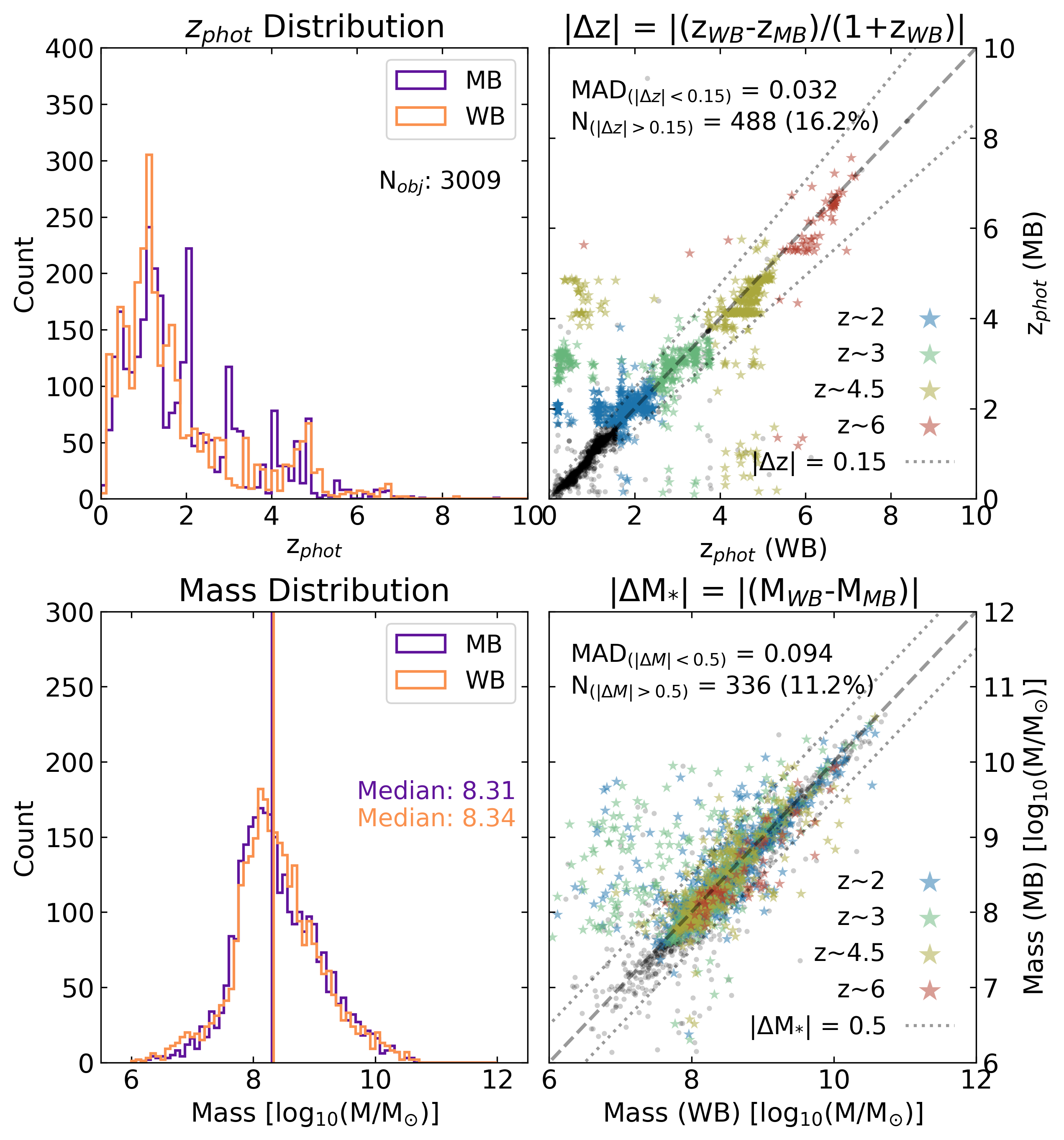}
\caption{\emph{Top}: Photo-z distributions (left) and scatter plot (right) for the wide band (WB; orange) and medium band (MB; purple) catalogs; \emph{Bottom}: Stellar mass distributions (left) and scatter plot (right) for the WB and MB catalogs. Galaxies with \OIIIHb \ emission  are shown as stars, and colour-coded by redshift bin.}
\label{fig_z_m_distributions_scatter}
\end{figure*}

\section{Data} \label{sec:data}

This work uses NIRCam imaging of the Canadian NIRISS Unbiased Cluster Survey (CANUCS, \citealt{Willott2022}) NIRCam Flanking field for MACS J0417.5-1154 (MACS0417 NCF), using five wide band (F090W, F115W, F150W, F277W, F444W) and nine medium band (F140M, F162M, F182M, F210M, F250M, F300M, F335M, F360M, F410M) photometric filters reaching 5-$\sigma$ depths for point sources of \tilda 29.3 and \tilda 29.0 mag for wide bands and medium bands respectively, supplemented with archival \emph{HST} WFC3/UVIS imaging of the same field in F438W and F606W to 5-$\sigma$ depths of \tilda 28.3 mag for point sources (Program ID 16667, PI M. Bradac). Together these provide continuous spectral coverage from 0.4-4.5$\mu$m, with every NIRCam medium band from 1.4-4.1$\mu$m. Gravitational lensing from the nearby galaxy cluster is negligible and is ignored in this analysis.

The NIRCam data were processed using a combination of the \emph{JWST} pipeline (version 1.8.0 and CRDS context jwst\_1001.pmap) and \texttt{grizli} version 1.6.0 \citep{Brammer_Matharu2021}. Both \emph{JWST} and \emph{HST} data were drizzled onto the same 0.04" pixel scale, and registered to Gaia DR3 astrometry. Further details concerning data reduction may be found in \cite{Noirot2023}. 

We obtained follow-up NIRSpec spectroscopy of MACS0417 NCF in multi-object spectroscopy (MOS) mode using the low-resolution prism (R\tilda100) over $\sim$0.6-5.3$\mu$m with \tilda3ks integration times. Further details of spectra reduction are provided in \cite{Withers2023}.

\section{Methodology} \label{sec:methodology}

\subsection{Photometry}\label{subsec:photometry}

The systematic differences of wide band (WB) vs. medium band (MB) photometry on photometric redshifts and stellar masses are investigated by creating parallel photometric catalogs from the full data set - one catalog which includes medium band photometry and one which does not. Point spread functions (PSFs) are determined empirically by median-stacking bright, non-saturated stars. We perform aperture photometry on PSF-convolved images homogenized to the F444W resolution using \texttt{SEP} \citep{Barbary2016}. For photometric redshift (photo-z) fitting, photometry is performed in fixed circular apertures of 0.3" diameter to enhance the S/N of the sample. Aperture corrections are applied to total fluxes based on the F444W PSF. Extinction is corrected using a \cite{Fitzpatrick1999} dust law with a selective extinction of R$_{v} =$ 3.1 and E(B-V) from the \cite{Schlafly2011} dust map. For SED-fitting, total fluxes are used to fit the overall normalization of the SED, and are measured in elliptical Kron-based apertures. Sources with circularized Kron radii $<$0.35" have total flux measured in fixed circular apertures of 0.7" diameter. 

\subsubsection{Synthesizing Wide Band Photometry From Medium Band Photometry}\label{subsubsec:synth_photometry}
The wide+medium band catalog (hereafter referred to as the ``medium band catalog") utilizes all 16 filters in the data set, whereas the wide band-only catalog uses the seven \emph{JWST} \& \emph{HST} wide band filters, leaving gaps in the spectral coverage at 2$\mu$m \& 3.5$\mu$m. To provide full wide band coverage out to 4.5$\mu$m we estimate photometry for F200W from F182M \& F210M and F356W from F335M \& F360M by synthesizing the total flux through the missing wide band based on measured total fluxes in the constituent medium bands, then dividing by the bandwidth of the synthesized bandpass to obtain flux density. Total flux in the missing wide band is synthesized as follows: (I) We compute the average flux per unit transmission element of the medium bandpasses, equal to the flux density multiplied by the filter bandwidth divided by the integral of the filter transmission curve; (II) The wide bandpass is split into three regions: blueward of the blue medium band pivot, redward of the red medium band pivot, and the central region between the two medium band pivot wavelengths; (III) Total flux in the blue region of the wide band being synthesized is equal to the average flux per unit transmission element of the blue medium band multiplied by the area under the transmission curve in the blue region of the wide band. The same is done redward of the red medium band pivot with the average flux per unit transmission element of the red medium band; (IV) Total flux in the central region is computed in the same manner, except the contributions from each filter are weighted by the relative transmissions of the two medium band transmissions at each wavelength.

\begin{figure*}[t!]
\plotone{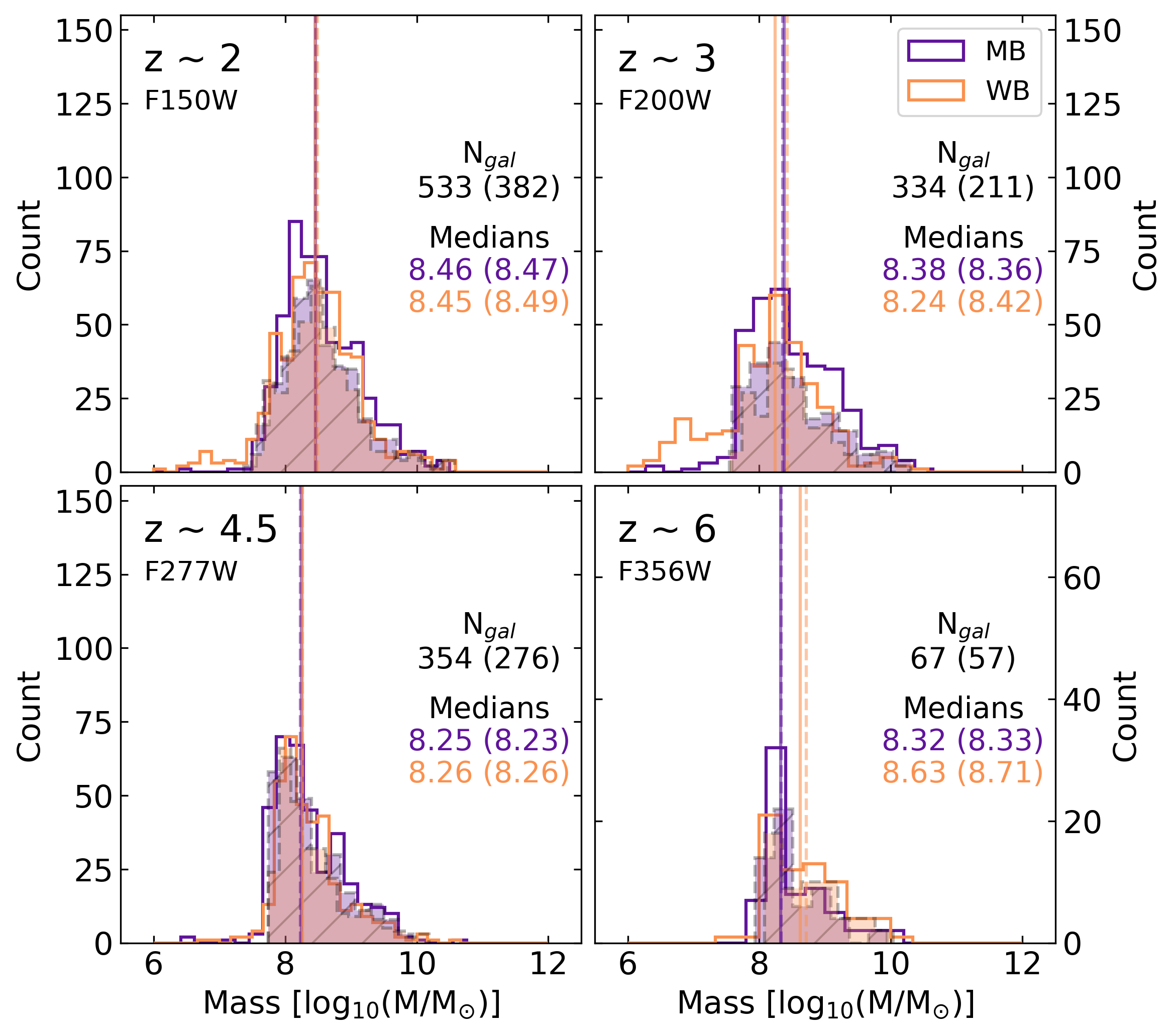}
\caption{Mass distribution of galaxies with strong \OIIIHb \ emission lines by redshift bin, with corresponding photometric wide band filter labelled. \emph{Clockwise from top left}: z\tilda2 (F150W), z\tilda3 (F200W), z\tilda4.5 (F277W), and z\tilda6 (F356W). The number of galaxies in each bin is shown along with the median stellar mass determined by wide (orange) and medium (purple) band photometry. Solid vertical lines show median values. The filled histograms, dashed lines, and values in parentheses shows only those galaxies in each bin with \deltaz\lessthan0.15.
}
\label{fig_sample_dist}
\end{figure*}

\subsubsection{Photometric Catalog Properties}\label{subsubsec:catalog_properties}
The full photometric catalog for the combined \emph{JWST}/\emph{HST} footprint contains 11,055 sources. Our science sample is limited to the overlap of NIRCam and WFC3/UVIS (6.9 arcmin$^2$, \tilda70\% of the NIRCam footprint), ensuring all galaxies have optical coverage. In order to not have our result dominated by low S/N photometry, we remove all sources with SNR \lessthan\ 10 in F277W. As the wide band catalog serves as the baseline for our comparison, we restrict our analysis to sources with redshifts out to z \tilda 10 and stellar masses (\Mstar) of $10^6$ $<$ \Mstar $<$ $10^{12}$ \Msol using wide band photometry. The final source catalog contains 3009 galaxies.

\subsection{Redshift- and SED-fitting}\label{subsec:sed_fitting}

Photometric redshifts are obtained with \texttt{EAZY} using 0.3" fixed-aperture photometry. The standard \texttt{EAZY} template set (\emph{tweak\_fsps\_QSF\_12\_v3}) is augmented with 3 templates from \cite{Larson2023} (Set 3) which incorporate synthetic \texttt{BPASS} spectra with \texttt{CLOUDY} nebular and continuum emission, reduced Lyman alpha emission (escape fraction $\sim$10\%), with stellar population ages of 6, 6.5, and 7 Gyr respectively. A 2\% systematic error and a magnitude prior are applied when fitting redshifts with \texttt{EAZY}. Note: as \texttt{EAZY} does not have magnitude prior templates for \emph{JWST} NIRCam filters at time of publication, the \emph{HST} WFC3/IR F160W prior was used and assigned to NIRCam F150W owing to the similarity of their wavelength coverage (1.272-1.808$\mu$m for F160W; 1.183-1.819$\mu$m for F150W).

SED-fitting is performed using \texttt{Dense Basis} \citep{Iyer2019} with total fluxes at the best-fitting \texttt{EAZY} redshift $\pm$0.01. The \texttt{Dense Basis} atlas, which provides a coarse mapping from the galaxy’s stellar population parameters to their corresponding SEDs, was generated with a flat star formation history prior, \cite{Calzetti2000} dust law, \cite{Chabrier2003} IMF, and an exponential reddening prior with scale value A\textsubscript{v} $=$ 0.33 mag.

Figure \ref{fig_z_m_distributions_scatter} shows photometric redshift (``photo-z"; top) and stellar mass (bottom) properties of the two catalogs, with distributions (left panel) and wide band vs. medium band scatter plots (right panel). We measure the degree to which photo-z's agree between the wide band and medium band catalogs using \deltaz \ = $|$z\textsubscript{WB}-z\textsubscript{MB}$|$/(1+z\textsubscript{WB})  $\leq$ 0.15, and find good agreement for 83.8\% of all sources. We measure agreement between galaxy stellar masses as \deltaM \ = $|$M\textsubscript{WB}-M\textsubscript{MB}$|$ \lessthan\ \ 0.5 dex, where stellar masses agree for 88.8\% of sources between catalogs.

\section{Results and Discussion}\label{sec:results}

Strong emission lines in high redshift galaxies are expected for those undergoing elevated rates of star formation. This proves challenging for SED-fitting codes with WB photometry alone as WBs may not distinguish between strong, discrete emission lines from young stellar populations at high redshift and generally more elevated continuum emission from older stellar populations at lower redshift, leading to an overestimation of stellar masses at z\tilda6 of up to 30\% \citep{Schaerer2009}.

The right-hand column of Figure \ref{fig_z_m_distributions_scatter} shows coloured stars for galaxies with \OIIIHb \ emission likely detected as an elevated flux within the wide or either of its constituent medium bandpasses, colour-coded by redshift bin. Bin widths are set by the redshift range corresponding to \OIIIHb \ within each wide band filter (e.g. \tilda 1.6 \lessthan \ z \lessthan \ \tilda 2.4 for F150W), for all wide bands from F150W to F356W. We find 1460 galaxies across all bins and visually inspect the best-fit SEDs, rejecting any galaxy whose SEDs do not show elevated flux likely due to \OIIIHb \ emission in either the wide or constituent medium bandpasses. Our final sample of candidate strong line emitters consists of 1288 galaxies across all bins, shown as stars in Fig. \ref{fig_z_m_distributions_scatter} and colour-coded by redshift bin.

Figure \ref{fig_sample_dist} shows stellar mass distributions for wide (orange) and medium (purple) band photometry in each redshift bin at z  \tilda2 (F150W), \tilda3 (F200W), \tilda4.5 (F277W), and \tilda6 (F356W). The filled orange and purple-hatched histograms show distributions for galaxies in each bin with \deltaz \lessthan 0.15. For the \tilda 70\% of galaxies where photo-z's agree between the wide and medium band catalogs, SED-fitting yields a median \deltaM \lessthan 0.2 dex across all redshifts. From SED visual inspections we note that many of the discrepant wide band redshift/SED-fitting solutions arise from the wrong Balmer line being fit in the relevant wide bandpass, or from low-redshift quiescent/dusty star forming galaxies. This results in median \deltaM \greaterthan 0.4 dex across all bins for the \tilda 30\% of galaxies with discrepant photo-z's.

\subsection{[OIII]/H\texorpdfstring{$\beta$}{b} \ with Wide versus Medium Band Photometry}\label{subsec:OIII_Hb}

\begin{figure}[t]
\plotone{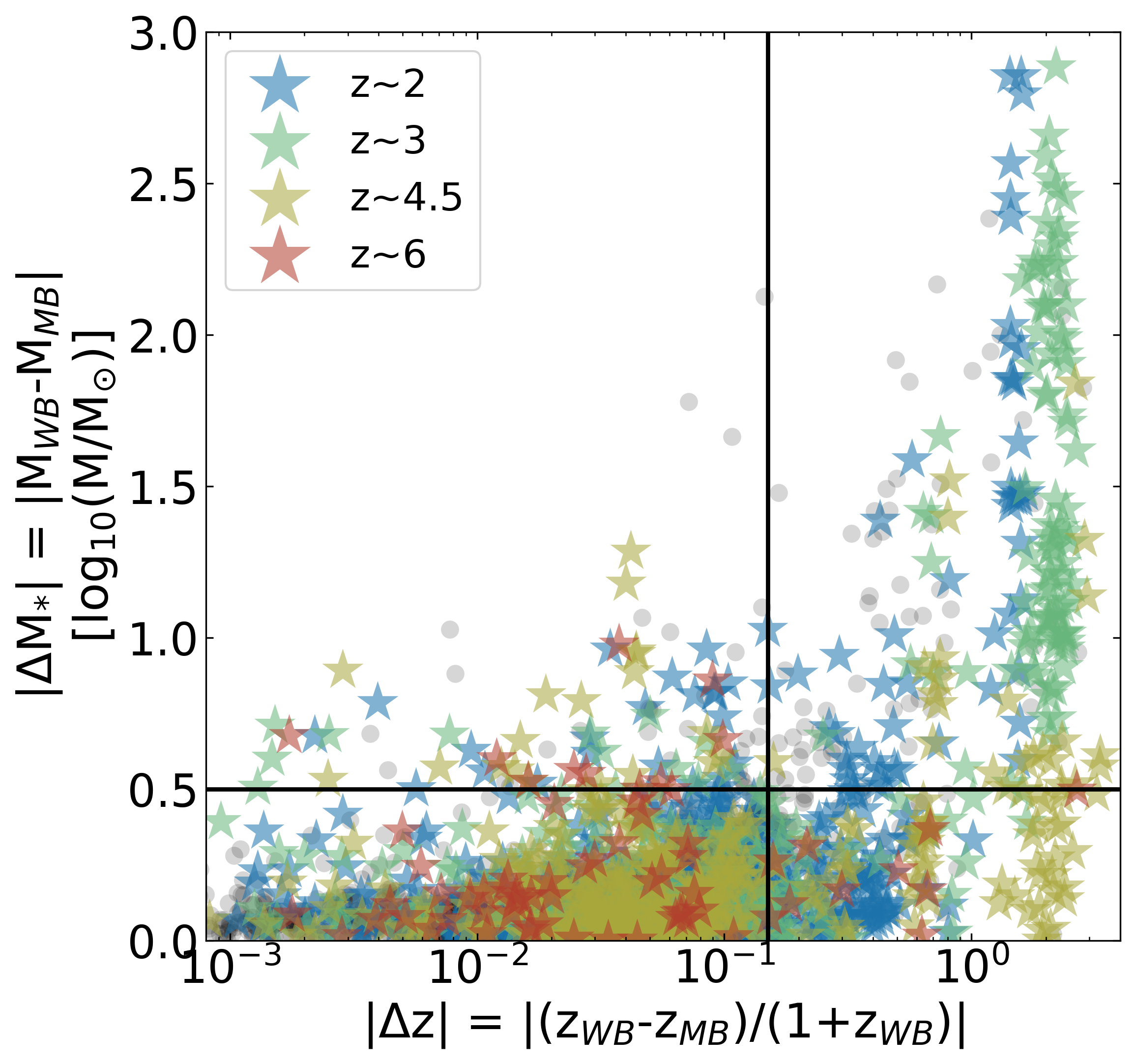}
\caption{Mass residuals versus photo-z residuals (abs. values). Boundaries demarcating mass and photo-z outliers are marked by solid black lines (\deltaM $>$ 0.5 dex, \deltaz $>$  0.15 respectively). Redshift bins are colour-coded as in Fig. \ref{fig_z_m_distributions_scatter}.
}
\label{fig_delm_v_delz}
\end{figure}

\begin{figure*}
\plotone{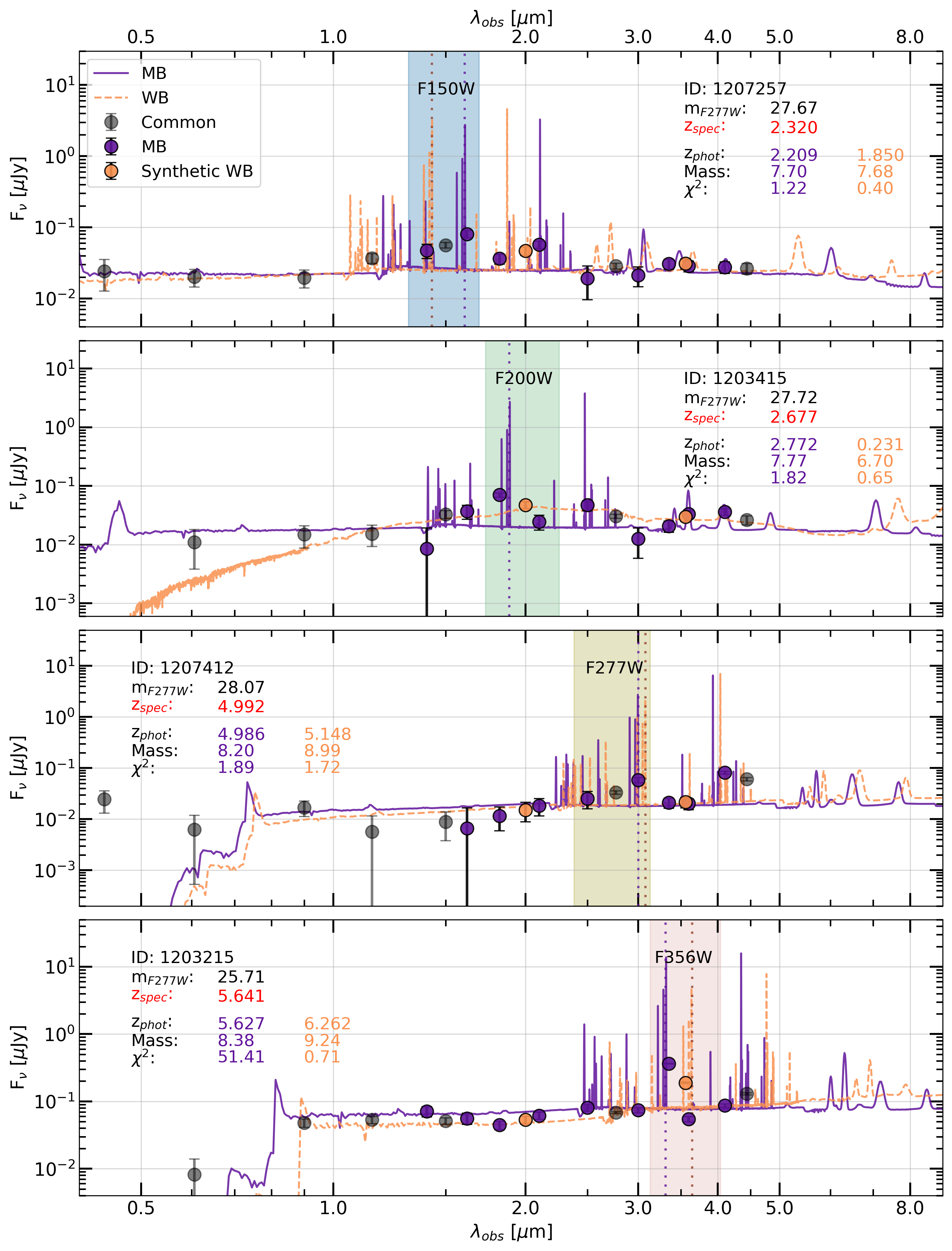}
\caption{\emph{Top to bottom}: Photometry and SEDs of sources with strong \OIIIHb \ emission in bins of ascending redshift. Wavelength coverage for each NIRCam wide bandpass is shaded, corresponding to the observed frame wavelengths of the \OIIIHb \ lines spanned by each redshift bin. Medium band photometry is shown in purple, synthetic wide band photometry in orange, and common photometric points in black. Catalog IDs, F277W magnitudes, and spectroscopic redshifts are listed for each source along with photometrically derived redshifts, masses, and $\chi$$^{2}$ from SED-fitting for both wide and medium band catalogs. The position of the \OIIIHb \ complex in each SED is marked by a vertical dotted line.}
\label{fig_SEDs}
\end{figure*}

Our final sample of strong line emitters contains 533, 334, 354, and 67 galaxies in bins at redshift \tilda2, \tilda3, \tilda4.5, and \tilda6 respectively, with 72\%, 63\%, 78\%, and 85\% of galaxies having consistent photo-z's between catalogs in each bin. When only considering galaxies with \deltaz \ \lessthan\ \ 0.15, the median stellar mass residual \deltaM \ = $|$M$_{WB}$-M$_{MB}|$ \ is \lessthan\ 0.2 dex across all bins, however the high fraction of galaxies placed at discrepant redshifts with wide band photometry (\tilda15-40\% depending on redshift bin) can bias these median stellar mass residuals within individual sample bins by 0.4-1.1 dex.

The treatment of the wide band filters for galaxies in our sample varies from galaxy to galaxy, but in general the \OIIIHb \ lines are either detected by the wide bands (and the photo-z's are likely to be in agreement) or not, and stellar masses derived at the fitted photometric redshifts either agree or not. 
Our expectation is that in cases where photo-z fitting yields similar redshifts, SED-fitting at those redshifts should yield similar stellar masses. Similarly, SEDs fit at different redshifts should in general yield inconsistent stellar mass estimates.

This is most clearly seen by considering mass residuals as a function of photo-z residuals, as in Figure \ref{fig_delm_v_delz} (all absolute values). Solid black lines at \deltaM $=$ 0.5 dex and \deltaz $=$ 0.15 split the residuals space into quadrants where masses and photo-z's agree between catalogs (to the left and below the lines, respectively) or do not agree (to the right and above the lines, respectively). Sources in our final sample are colour-coded by redshift bin as in Fig. \ref{fig_z_m_distributions_scatter}.

In the lower left quadrant of Fig. \ref{fig_delm_v_delz} are galaxies for which photo-z's and stellar masses agree between both catalogs (80.4\% of all 3009 galaxies in this study; 66.8\% of all 1288 galaxies in our final sample), in line with our initial expectation.

\subsubsection{Redshifts Agree, Masses Disagree}\label{subsubsec:same_z_diff_m}

Of the galaxies in our final sample we find 66 cases (\tilda5\%) where stellar masses between catalogs disagree despite similar redshifts (top left quadrant in Fig. \ref{fig_delm_v_delz}). In these cases, owing to the coarser sampling of the SED, wide band photometry distinguishes the emission lines from continuum, but underestimates the strength of the lines themselves, inferring lower instantaneous star formation rates and older stellar populations, resulting in higher WB stellar masses. This behaviour is more pronounced at higher redshift, with 15\% of galaxies in the z\tilda6 bin having discrepant masses despite having consistent photo-z's, resulting in \deltaM \tilda0.6 dex across all bins.

\subsubsection{Redshifts Disagree, Masses Disagree}\label{subsubsec:diff_z_diff_m}

For 7.7\% of all galaxies in our study (233 out of 3009) we find that discrepant redshifts lead to discrepant masses. Of these galaxies, 71\% are included in our final sample (top right quadrant in Fig. \ref{fig_delm_v_delz}). In the cases where the \OIIIHb \ emission is not detected in the wide band SED, it is confused either for another Balmer line or continuum emission from older stellar populations at lower redshift. This generally results in discrepant photo-z's, which in turn yields discrepant stellar masses between the wide and medium band catalogs, with a median \deltaM \ across all bins of \tilda1.2 dex.

\subsubsection{Redshifts Disagree, Masses Agree}\label{subsubsec:diff_z_same_m}

We note degenerate cases where stellar masses do not change significantly despite a discrepancy in photo-z's in 8.5\% of all galaxies (255 of all 3009 galaxies), of which \tilda75\% are included in our final sample, mostly in the z\tilda2 \& z\tilda4.5 bins (lower right quadrant in Fig. \ref{fig_delm_v_delz}, blue and gold stars). These sources are being placed at the wrong cosmological epoch, artificially decreasing the stellar mass density at earlier times while simultaneously contaminating the low-mass quiescent galaxy population at lower redshifts. 

These degenerate cases are represented in the high fraction of galaxies with \deltaz \ \greaterthan \ 0.15 (\tilda 30\% of our final sample across all bins), and we speculate that the discrepancy in photo-z's may be mitigated by better observed frame optical coverage at $\lambda$ $<$ 1$\mu$m to rule out lower-redshift solutions during photo-z fitting. Confirmation of this hypothesis, and the degree to which it affects derived galaxy properties, we leave to future work in the CANUCS Frontier Fields NCF.

\subsubsection{Spectral Energy Distributions}\label{subsubsec:SEDs}

Figure \ref{fig_SEDs} presents examples of wide and medium band SEDs of galaxies for which we obtained spectroscopic redshifts in each bin, with the observed frame wavelengths corresponding to the redshift range of the bin shaded in each panel, and labelled with the corresponding wide bandpass. Medium band photometry is shown in purple, synthetic wide band fluxes in orange, and wide bands common to both catalogs are in black. Vertical dotted lines mark the position of the \OIIIHb \ lines for each SED (when detected). 43 redshifts have been confirmed with NIRSpec follow up spectroscopy (see \citealt{Withers2023} for details of spectroscopic data reduction), with 10-11 redshifts per redshift bin. The galaxy SEDs presented in Fig. \ref{fig_SEDs} are representative examples reflecting the characteristic behaviour of wide versus medium band photometry shown in Fig. \ref{fig_delm_v_delz} and discussed in \S\ref{subsec:OIII_Hb}, and the availability of spectroscopic redshifts allows us to better evaluate the quality of photometric measurements made with both filter sets. In all cases the MB photo-z is closer to the spectroscopic redshift than the WB photo-z.

The first panel (ID 1207257) shows \OIIIHb \ detected in both the wide and medium bands with similar stellar masses, however the wide band photo-z is nearly a catastrophic failure ($|$z$_{spec}$-z$_{phot}$$|$/(1+z$_{spec}$)$|$ $=$ 0.14), demonstrating the power of the medium bands to accurately constrain photo-z's. The second panel (ID 1203415) shows strong \OIIIHb \ emission in F182M not being detected in the wide bands, and is confused for a low-redshift quiescent galaxy. The confusion between high-z star forming galaxies and low-z dusty/quiescent galaxies is most common in the z\tilda3 bin (see top right quadrant of Fig. \ref{fig_delm_v_delz}). The last two panels (1207412 \& ID 1203215) show \OIIIHb \ detected in both wide and medium bands with consistent photo-z's, however the emission line is placed at the wrong position within the wide bandpass and the strength of the emission lines are underestimated due to the coarser sampling of the SED, as discussed in \S\ref{subsubsec:same_z_diff_m} (top left quadrant of Fig. \ref{fig_delm_v_delz}). This occurs in \tilda5\% of all galaxies in our sample across all bins, and drives a median \deltaM \ of \tilda0.7 dex, demonstrating that reasonably accurate photo-z's do not necessarily yield reliable stellar mass estimates.

\begin{figure}[t]
\plotone{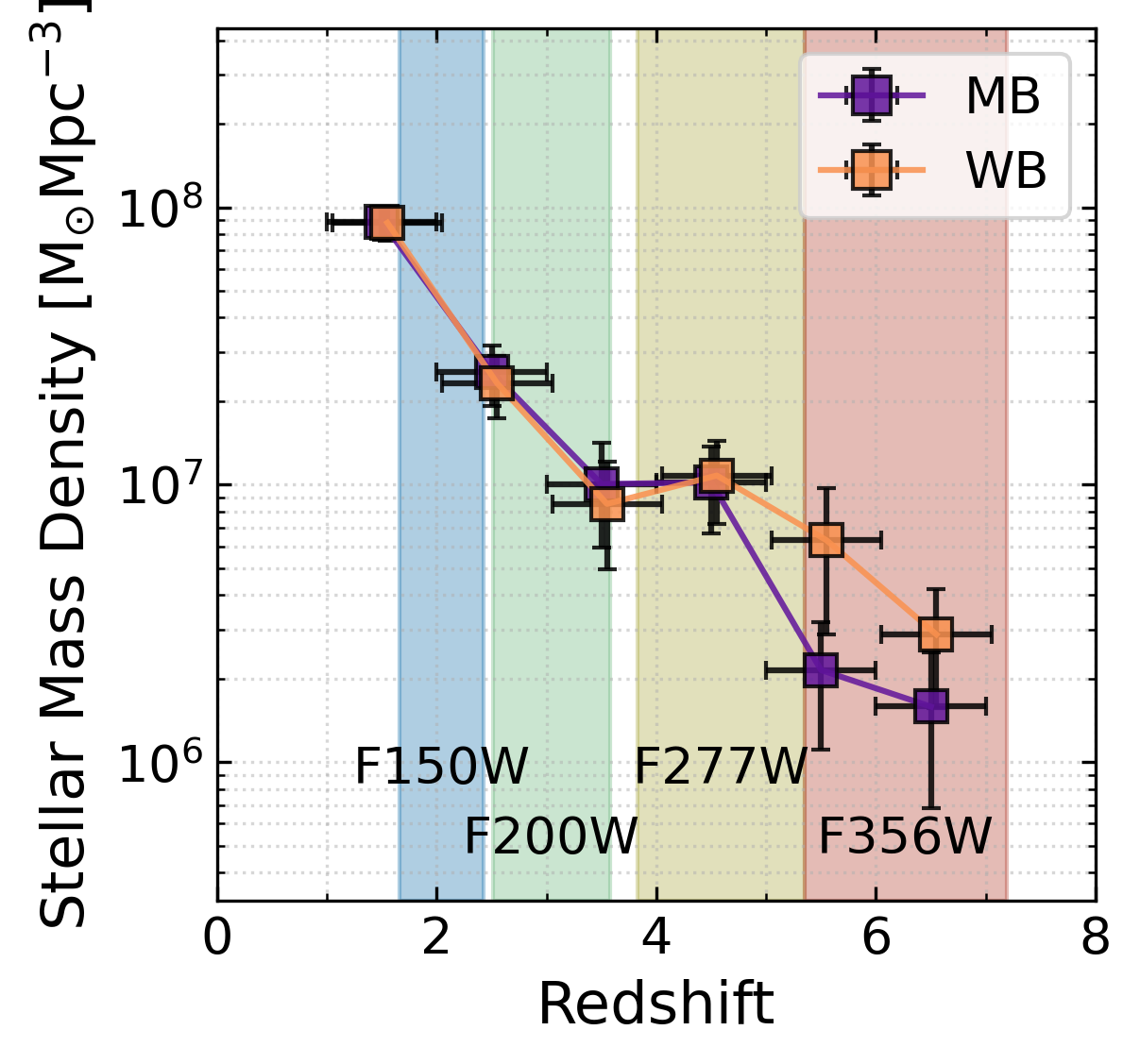}
\caption{Stellar mass density for both wide and medium band catalogs as a function of redshift. NOTE: SMD based on a single NIRCam pointing covering \tilda7 arcmin$^2$ and does not account for cosmic variance (affecting both the overall normalization and error budget), however the ratio of the SMD between catalogs is the salient quantity. Shaded regions denote the extent of each redshift bin.
}
\label{fig_SMD}
\end{figure}

\subsection{Cumulative Effect on Stellar Mass Density}\label{subsec:SMD}

We have shown how medium band photometry can yield significantly different photometric redshifts and galaxy stellar masses as compared to wide band photometry for galaxies beyond z \tilda \ 2. Acknowledging the limited statistics of our sample and the large uncertainty due to cosmic variance, which will not capture the high-mass end of the stellar mass function, and the modest supporting data at $\lambda$ \lessthan\ \ 1$\mu$m, we nevertheless consider the impact this early result will have on a benchmark observable diagnostic of galaxy evolution: the stellar mass density (SMD). While our SMD in this single field is surely underestimated for both WB's and MB's, our primary interest is not in the absolute value of the SMD for either photometry set but rather the ratio of their SMDs, in order to investigate potential \emph{systematics} in the SMD revealed by the inclusion of medium band photometry.

Figure \ref{fig_SMD} shows the evolution of the SMD for the two catalogs as a function of redshift for MACS0417NCF. The results presented above show most \OIIIHb \ emitters move from low redshifts in the wide band catalog to higher redshifts in the medium bands, and we might naively expect the medium band SMD to be greater than that of the wide bands at high redshift and lower than the wide band SMD at low redshift. However, of these \OIIIHb \ emitters, the galaxies with discrepant stellar masses mostly have \Mstar \lessthan\ $10^{8}$\Msol \ (see lower-right panel of Fig. \ref{fig_z_m_distributions_scatter}) and their contribution to the SMD is negligible compared with the abundance of more massive galaxies at lower redshifts. 

The two principal behaviours seen with wide band photometry presented in \S\ref{subsec:OIII_Hb} are: (I) strongly star forming galaxies at \tilda1.5 \lessthan\ z \lessthan\ \tilda3.5 are not fit at the appropriate cosmological epoch, reducing the SMD below z\tilda3.5; and (II) star formation rates can be underestimated and older stellar populations inferred at higher redshifts, artificially inflating the SMD at earlier times. This effect is most prominently seen in the highest redshift bin (red-shaded region in Fig. \ref{fig_SMD}; bottom panel in Fig. \ref{fig_SEDs}).

As a consequence of these two effects, the overall SMD is 10-20\% higher in the medium bands at z \lessthan\ \ 4, but at z \greaterthan \ 5 the opposite is true and the wide band photometry begins to systematically overestimate stellar masses, driving a SMD that is up to 2-3 times higher than that measured with medium band photometry. The ratio increases significantly beyond z \tilda5, consistent with the literature (\citealt{Schaerer2009, Stark2013, Smit2014}), and indeed may help reconcile the high number density of massive, high redshift galaxies found in recent \emph{JWST} surveys (e.g. \citealt{Labbe2023}).

\section{Conclusion}\label{sec:conclusion}

In this paper, we have introduced an analysis of systematic effects which arise between data sets employing wide band photometry alone, versus a combination of wide and medium bands. We find that:
\begin{itemize}
    \item Elevated star formation beyond z\greaterthan2 compromises the ability of wide band photometric filters to detect strong \OIIIHb \ emission lines at $\lambda$\greaterthan 1.5$\mu$m (observed), with failure rates of: 28\% at z\tilda2, 37\% at z\tilda3, 22\% at z\tilda4.5, and 15\% at z\tilda6. The median \deltaM \ for such cases is \tilda0.4 dex.
    \item For galaxies where photo-z's agree between wide and medium band photometry, the \deltaM is \lessthan0.2 dex across all redshifts/filters. 
    \item Wide band photometry can underestimate star formation rates and overestimate stellar masses beyond z\tilda5, even when the photo-z agrees with the medium band-derived redshift. This occurred for 15\% of our strong line emitters at z\greaterthan5, driving a median \deltaM \ of \tilda0.7 dex. 
    \item The stellar mass density of the universe is potentially overestimated with wide band photometry by up to a factor of 2-3 at redshifts  z \greaterthan\ 4, consistent with previous studies, and potentially underestimated at z \lessthan\ 4 by 10-20\%.
\end{itemize} 

We argue that inclusion of medium band photometry is integral to detecting strong emission lines beyond 1.5$\mu$m, properly constraining galaxy SEDs and obtaining accurate physical properties, and should be included in the design of photometric surveys with \emph{JWST} targeting high-redshift galaxies, in particular for the epoch at z\greaterthan5 when such lines become ubiquitous in galaxy SEDs.

\begin{acknowledgments}

This research was enabled by grant 18JWST-GTO1 from the Canadian Space Agency, and funding from the Natural Sciences and Engineering Research Council of Canada. 
YA is supported by a Research Fellowship for Young Scientists from the Japan Society of the Promotion of Science (JSPS). MB acknowledges support from the Slovenian national research agency ARRS through grants N1-0238, P1-0188 and the program HST-GO-16667, provided through a grant from the STScI under NASA contract NAS5-26555.
This research used the Canadian Advanced Network For Astronomy Research (CANFAR) operated in partnership by the Canadian Astronomy Data Centre and The Digital Research Alliance of Canada with support from the National Research Council of Canada the Canadian Space Agency, CANARIE and the Canadian Foundation for Innovation.

\end{acknowledgments}

\facilities{JWST, HST}

\software{
          EAZY \citep{Brammer2008},
          Dense Basis \citep{Iyer2019},
          SEP \citep{Barbary2016},
          photutils \citep{larry_bradley_2023_7946442},
          astropy \citep{astropy:2022}
          Source Extractor \citep{BERTINE.ARNOUTS1996}
          }

\bibliography{MB_letter_v2_OIII_Hb}
\bibliographystyle{aasjournal}

\end{document}